# NASA Astrophysics Division Research and Analysis Program Statistics for the Period 2001-2009

Eric P. Smith
*NASA Headquarters Science Mission Directorate, Astrophysics Division*

## Abstract

I describe the various elements of the NASA Science Mission Directorate's Astrophysics Division Research and Analysis Program and provide quantitative descriptions for factors such as proposal submission characteristics, proposal success rates, distribution of science areas for selected proposals, as well as funding distributions for the various program elements. I examine the variation of these factors with time to explore possible trends. The measures described here can be used as starting points for future discussions about issues related to balance within the astronomy and astrophysics research and analysis program.

# 1. Introduction

## 1.1. NASA Astrophysics Research and Analysis Programs

NASA funds a wide variety of astrophysics and fundamental physics research efforts through several different means. Direct solicitations for proposals issued from NASA Headquarters, solicitations issued by organizations operating NASA's major astronomy observatories (e.g., the *Hubble Space Telescope* or the *Chandra X-Ray Observatory*) and through grants to scientists serving on mission definition teams and science working groups. Responses to these solicitations are peer-reviewed to help NASA select the research projects. In fiscal year 2009 the NASA Astrophysics Division spent approximately $230M on such efforts. For the purposes of this study I will refer to the collective science covered by these programs as 'astrophysics' for economy of notation, but with the understanding that this includes astronomy, astrophysics and fundamental physics research.

This study is restricted to only those elements of the research and analysis programs (R&A) that are solicited yearly through the annual NASA headquarters Research Opportunities in Space and Earth Sciences (ROSES[1]) call, and are aimed at a wide audience of scientists. I examine those research programs from the years 2001 through 2009. I do not consider here those individual research investigations associated with operating observatories both large (*Hubble*, *Chandra*, *Spitzer*) and small (*e.g.*, *Swift*, GALEX). I will concentrate this examination to the following elements of the ROSES: the Astrophysics Data Program (ADP), the Astronomy and Physics Research and Analysis Program (APRA) and the Astrophysics Theory Program (ATP).

The science community and NASA recognize the value in maintaining a reasonable balance across the various disciplines and topics covered by the NASA astrophysics research program. This balance is difficult to define precisely because it is widely

---

[1] Research Opportunities in Space and Earth Science 2010 available online at http://nspires.nasaprs.com/external/solicitations/summary.do?method=init&solId={96364 73D-602B-F49F-ABDC-5A26F36D08CD}&path=open

recognized that some areas of research are inherently more costly than others to perform (*e.g.*, a suborbital balloon payload development program vs. a theory grant) and some areas of astronomy have far more practitioners than others. Moreover, the very nature of scientific inquiry means that some areas of research decrease in importance or relevance with time while new avenues open up. Thus, balance cannot be simply defined as equal amount of funding per selected proposal, equal numbers of grants per topic, or constancy of funding over time. Here, I define balance to mean only a distribution of grants and funding that strictly reflects proposal pressure and scientific judgment arising from the peer-review process. That is, a program element is defined as being "in balance" when the fraction of the selected program equals the fraction of the proposal pressure for that element. A more global definition of program balance might take into account the inherent financial weighting of some efforts and NASA-specific concerns.

In the following sections I give a brief description of each of the research programs. Section 2 provides details about the data used in the study. In section 3 I describe various characteristics of the data and examine it for any discernable trends. Finally, section 4 discusses possible reasons behind some of the features seen in the data.

### 1.1.1. Astrophysics Data Analysis Program (ADP)

The ADP solicits research involving NASA space astrophysics data that are currently archived in the public domain at the time of proposal submission. Most of these data have undergone considerable reduction and refinement by way of calibrations and ordering, and extensive data analysis software tools often exist for these data. At present, scientists

- 3 -

can propose for research funding to analyze data from 32 missions covered by this program. The ADP represents NASA's investment in maximizing the scientific benefit of its operating or past missions. Some operating missions conduct their own archival research programs (*e.g.*, *Hubble*, *Chandra, Fermi*) for NASA. Submissions to those programs are not considered in subsequent analyses here. Over time, as the number of archival datasets increases the importance of this program will certainly grow. In addition to examining proposals submitted to the ADP program I have included submissions to the Long Term Space Astrophysics Research Program (LTSA). This program, discontinued in 2005, funded researchers for programs similar to those associated with the current ADP. NASA has made recent changes in the duration of the ADP awards in order to recapture efforts of similar scale to those that were previously funded by the LTSA program. To that end, the present ADP supports research efforts up to four years in duration.

When submitting an ADP proposal PIs are asked to select one of the 10 Research Areas defined for the program. Full definitions of the areas are given in ROSES-2009. Here we simply list the topic titles:
- Star Formation and Pre-Main Sequence Stars
- Stellar Astrophysics
- Post-Main Sequence stars and Collapsed Objects
- Binary Systems
- Interstellar Medium and Galactic Structure
- Normal Galaxies
- Active Galaxies and Quasars
- Large Scale Cosmic Structure

- Atomic/Molecular Databases
- Other

I shall use these topics when assessing balance across areas of scientific investigations within the ADP.  NASA funded 126 new and continuing ADP investigations for a total cost of $11.9M in government fiscal year 2009.

### 1.1.2. Astronomy and Physics Research and Analysis Program (APRA)

The Astrophysics Research and Analysis program solicits basic research proposals for investigations that are relevant to NASA's programs in astronomy and includes research over the entire range of photons, gravitational waves, and particles of cosmic origin.

The APRA program seeks to support research that addresses the best possible (i) state-of-the-art detector technology development for instruments that may be proposed as candidate experiments for future space flight opportunities; (ii) science and/or technology investigations that can be carried out with instruments flown on suborbital sounding rockets, stratospheric balloons, or other platforms; and (iii) supporting technology, laboratory research, and/or (with restrictions) ground-based observations that are directly applicable to space astrophysics missions. To meet these goals, proposals are solicited in the following five broad categories:

- Detector Development
- Suborbital Investigations (balloons and sounding rockets)
- Supporting Technology
- Laboratory Astrophysics
- Ground-Based Observations.

In addition to categorizing the proposals based upon those areas, NASA additionally categorizes the proposals into science disciplines largely based on the differing technologies associated with each:

- Gamma-ray Astrophysics
- X-Ray Astrophysics
- Ultraviolet/Optical Astrophysics
- Infrared/Submillimeter and Radio Astrophysics
- Particle Astrophysics
- Laboratory Astrophysics

NASA funded 188 new and continuing APRA investigations (including both principal investigator and co-investigator proposals) for a total cost of $44M in government fiscal year 2009.

### 1.1.3. Astrophysics Theory Program (ATP)

The Astrophysics Theory program (ATP) supports efforts to develop the basic theory for NASA's space astrophysics programs. Theoretical proposals submitted for this program must both: be directly relevant to space astrophysics goals by facilitating the interpretation of data from space astrophysics missions or by leading to predictions that can be tested with space astrophysics observations; and consist predominantly of theoretical studies and the development of theoretical models.

Proposals submitted to the ATP must include a PI selected Topic Category, analogous to the ADP Research Areas. Titles for these are:

- Star Formation
- Stellar Astrophysics
- Collapsed Objects and X-ray Astrophysics
- Supernovae and Gamma-ray Bursts
- Interstellar Medium, Cosmic Rays and Galactic Structure
- Normal Galaxies
- Active Galaxies and AGNs
- Large Scale Structure and Dark Matter
- Dark Energy and the Cosmic Microwave Background
- Gravitational Astronomy
- Other

NASA funded 117 new and continuing ATP investigations for a total cost of $13.7M in government fiscal year 2009.

## 2. Data Assembly

### 2.1. Data Sources

NASA maintains a database of proposal submission information for each of the programs under consideration. Data for the years 2001-2009 are homogeneous and nearly uniform in quality and completeness. Prior to 2001 the research solicited in the three programs discussed above was tracked in a very different manner and in several additional programs. For example, the current APRA program elements were divided into separately competed wavelength specific programs. Also, data for FY2000 and earlier were much less complete for all the data fields needed to track research areas, and awarded funding amounts. Prior to 2001 programs tracked their awarded amounts in separate ways, at the discretion of the individual program officers. Table 1 gives the information tracked for each proposal in this study.

## 2.2. Data Preparation

While the data were largely uniform as retrieved from the NASA Peer Review Services archives, certain information was occasionally missing for individual proposals, or for specific years. For example, the linked organization information has become uniform and reliable after 2007. I constructed pre-2007 linked organization data from the Principal Investigator (PI) host institution name in conjunction with the associated PI. In most cases the PI host institution and the linked organization are identical. However important exceptions occur in situations where individuals are employed by not-for-profit organizations but hosted at a NASA center or where the affiliation of the individual is not clear simply from the host institution name (*e.g.*, a California Institute of Technology employee may be a faculty member in Pasadena or a Jet Propulsion Laboratory employee). Five individuals did not have linked organizations listed in the Central Contractor Registration database. These individuals were either self-employed or worked for organizations that did not register with the government. The information from these individuals was used for all subsequent analyses except those related to institutional performance. The small number of such cases will not alter subsequent findings.

Proposers enter the research area or discipline area data at the time of proposal submission. These fields are optional however and are occasionally left blank. I entered appropriate choices for those cases where the proposer left the field blank into the database based upon information in the proposal's title and research summary fields. I made no attempt to verify the appropriateness of the research area or discipline area data entered by proposers.

In some instances proposals are selected for partial funding. All these proposals were flagged as 'Accepted' even though some of the partial funding was clearly intended as bridge funding and could not support the full research program proposed. The rationale for counting these as 'Accepted' proposals was that NASA was devoting funding to that institution/research area/individual and this was a quantity that I was interested in tracking.

## 3. Results

### 3.1. Proposal Pressure and Program Funding History

Proposal pressure, or the relative abundance of any particular research area or discipline topic in a given cycle of proposal submissions, is a common way to discern scientific trends. NASA accepts proposals for space astronomy across a wide variety of areas and disciplines in every proposal cycle. Those areas of research that are perceived to be "hot topics" attract larger shares of proposals than other areas. Measuring the distribution of proposals submitted in each of the research areas, and their variation with time could be used to track the ebb and flow of interest in any particular area. Tables 2-4 give the basic statistics about number of proposals submitted and accepted for each of the programs under consideration. Gaps and additional proposal cycles in the data for certain years arise from two sources. In 2005 the ADP was cancelled due to funding limitations. No new selections were made. In calendar 2006 there was an additional APRA call for proposals that is labeled as occurring in fiscal year 2006.5 throughout the study

There are some small, but noticeable shifts in research area emphasis over the period covered. In particular, for the ADP and ATP programs there is an increase in the number of proposals related to Dark Energy and gamma ray burst investigations. This increase is coincident with the greater community interest in Dark Energy in recent years and the launch of the *Swift* and *Fermi Gamma-Ray Space Telescope* missions. There is an apparent decrease in submissions in the category of Interstellar Medium and Galactic Structure in both the ADP and ATP programs. For the APRA program there has been a slight increase in the proposal pressure for the infrared/submillimeter/radio research area, and a slight decrease in cosmic-ray/particle astrophysics.

When discussing the funding history for the research and analysis (R&A) program it is instructive to place it in context with the rest of the NASA Astrophysics Division (AD) budget. I show in Figure **1** the proportional AD expenditures divided into three broad categories, mission development, mission operation, and research funding for the past decade. Here research funding represents the broad set of funds sent to the science community as described in §1.1, but without the funding associated with mission science teams. That component of the NASA funding for scientists is less than 10% of the total sent to the research community) For the past decade the fraction of funds distributed to the science community through ADP, APRA, ATP, and for investigations associated with observations from satellites has remained nearly constant at 11±0.09% of the AD budget per year. The fraction of astrophysics funding allocated to the ADP, ATP and APRA programs alone in those same years was 6±0.08% per year.

The relative decrease in mission development funds following FY2008 is due to several factors; a decreasing Astrophysics Division total budget, combined with an uncertainty of which major mission to move into development until the 2010 National Academy of Sciences astronomy and astrophysics decadal survey selected a clear community priority, and large number of missions continuing operations after their prime mission phase.

To show the relative sizes of the ADP, APRA, and ATP programs I plot in Figure **2** the amounts of new award first year funding for each of those programs from 2001 to 2009. The cost of a typical APRA grant is 2-3 times that of a grant for data analysis (ADP) or theoretical study (ATP) because of the costs associated with hardware development and the often larger teams involved.

The cyclical behavior of APRA funding level for first year awards reflects the combination of longer-term grants (up to five years) and larger award sizes associated with some of the efforts. If more than one of these efforts completes in a given year additional funds are available in subsequent years for new selections. The large increase from 2007 to 2008 for ADP is attributable to the inclusion of *Spitzer Space Telescope* data in the eligible archive pool. Funds previously held in the *Spitzer* program were reallocated in part to ADP in anticipation of this rise in demand. Funding for the ATP program has increased throughout the period, as recommended in the *Astronomy and Astrophysics for the New Millennium* report, but the oversubscription rate has not decreased.

## 3.2. Research Area Balance

As mentioned in §1.1 the notion of balance is important and can be examined with the information found in Tables 2-4. For each of the research areas or disciplines we note in Tables 2-4 the eight-year median for proposals received and accepted per year and from that calculate the fraction of proposals and acceptances in each area. The proposal fraction measures the proposal pressure and the acceptances would track them precisely if this pressure were the only consideration in making selections. Unlike the National Science Foundation however whose mandate is solely to do the best peer-reviewed science, NASA must take other factors into consideration when making selections. In some years proposals for specific topics may be highlighted within the annual *ROSES* call as being particularly sought by NASA, thereby influencing the balance of submissions. Also, retaining core competency at NASA centers in technical areas widely used by the astronomy community is one such consideration. It is clear from Table 5 however that the median proposal pressure and median accepted program fraction are, within the standard deviations, identical.

The information contained in Tables 2 through 4 can be graphically depicted as in Figures 3-6. Here, for each program year I represent the various elements of the program as stacked vectors. When the selected, or accepted, fraction of the program (number of selected investigations for a particular category as a fraction of the total number of selections) equals the incoming proposal pressure (number of proposed investigations for a particular category as a fraction of the total number of proposals for a particular program) for that element the vectors are horizontal, representing a "balance" of selection and pressure. Where the proposal pressure was greater than the selected fraction the left hand side of the vector is below the right hand side. Where the selected fraction exceeds the proposal pressure the right hand side is below the left hand side. The length of the vectors is proportional to the fractional proposal pressure for that research area element (i.e., longer vectors indicate a larger fraction of the proposals received that cycle were for that particular category within the program). These figures can allow one to quickly and visually assess the program balance and any trends with time.

## 3.3. Institutional and Individual Performances

Using the proposal submission data we can explore the tendencies of individuals and institutions that respond to the NASA solicitations. Demographic measurements such as the fraction of people submitting multiple proposals and the success measures for institutions and individuals are easily constructed. The number of proposals submitted per principal investigator (PI) can be a measure of both the difficulty of getting proposals accepted (presumably increasing the number of proposals per individual) and/or the collective creativity of the community (people with many ideas tend to submit many proposals). Institutional measures of success are of interest because they highlight those institutions that are particularly effective at winning NASA grants for whatever reasons and identify less successful institutions so that they might undertake examinations of how to improve their NASA proposal success rates.

Figures 7-9 show the distributions for number of proposals per PI submitted to each of the programs over the 2001 to 2009 period. In all cases the most common instance, or modal value, is for individuals to have submitted a single proposal. There are very prolific proposal writers however for each of the programs. For example, a single PI submitted 21 proposals to the ADP/LTSA programs over the study period. These figures also show that the frequent proposers, defined as PIs who submit on average more than one proposal every two years, experience a wide range of success rates. Frequent submission of proposals to any of the programs does not appear to be a guarantee of success.

Institutions which submitted more than 5 proposals over the 2001-2009 period and whose proposal acceptance rates were greater than or equal to the average acceptance rate for the individual programs are listed in Tables 2-4. NASA institutions had a slightly higher success rate (33%; defined as total number of accepted NASA institution proposals divided by the total number of NASA institution proposals submitted) than non-NASA institutions (28%). This difference arises largely from the performance in the ADP and APRA programs which are closely tied to NASA missions or spaceflight hardware development. There is no significant difference in the NASA and non-NASA acceptance rates for the ATP program.

## 3.4. Research Costs Comparisons

NASA uses scientific peer review to evaluate each of the proposals received in these three programs for three factors: scientific merit, relevance to NASA and cost reasonableness. Trends in the area of scientific merit are possibly reflected in the slow variation of program balance with time. There are no discernable trends in criteria of NASA relevance. Very few proposals that are ultimately peer reviewed are deemed not relevant to NASA. NASA instructs its peer review panels to evaluate the cost reasonableness of the proposed effort – do the proposed resources and skill mix match the scope of the work - as opposed to the cost itself. This is done because some institutions are inherently more expensive (*i.e.*, they have higher institutional overheads) and the individual proposer has no control over that fact. To measure whether proposed costs have any effect on a proposal's selection I compare the distributions of proposed year one and total costs of selected and non-selected proposals for each of the programs. For ease of comparison all costs were converted to fiscal year 2010 dollars using the NASA new start inflation indices. Table 5 displays the results of using the two-sided Kolmogorov-Smirnov test for each program to compare the distribution of proposed costs for selected and non-selected proposals. Figures 10-13 depict the distributions of proposal costs for both accepted [filled histograms] and rejected proposals. Distributions for both the total costs and year one only costs are also shown. The cumulative distributions of the costs (total and year one) for both accepted and rejected used to construct the Kolmogorov-Smirnov statistic *D* are also given. I compared both total costs and single year costs to remove any effects related to the length of the proposed effort affecting its 'selectability' based on cost.

There is no statistically meaningful difference in the distribution of proposed costs for accepted or rejected proposals in the ADP and ATP programs over the 2001-2009 interval. However, for the APRA program, the distributions of accepted and rejected proposal costs are statistically different at better than the 99% confidence level. Interestingly, the accepted proposals are, on average, slightly higher in cost. This is due to the inclusion of the more costly suborbital proposals that NASA funds through this APRA. Removing those proposals from consideration we see that there is no statistically significant difference between the cost distributions for accepted and rejected APRA proposals. In the next section I discuss why inclusion of Suborbital Investigations in the APRA comparison of cost might lead one to incorrectly conclude that more costly proposals are favored at a statistically significant rate.

## 4. Discussion

A clear lesson learned from examination of these data is the remarkable stability of the various facets of the R&A programs considered here. Over the 2001-2009 interval the relative constancy of funding proportion compared with the rest of the astrophysics portfolio shows that managers have understood the advice from the community regarding the importance of a vibrant basic research program (2000 and 2010 decadal surveys, recent NRC reports). Though there have been years when funding for specific elements of these programs have been significantly affected (cancellation of ADP in 2005, removal of the LTSA program altogether after 2004), these have been partially offset by increases in other areas of basic research (the additional APRA call in 2006 and longer award durations covered in ADP partially compensating for the loss of LTSA). It is important to note that all this is true in a proportional sense. With the currently projected declining astrophysics budget (FY11 President's budget proposal) NASA officials are faced with questions concerning the absolute level of funding for these important programs. It is not obvious that maintaining the historical fraction of the Astrophysics Division budget devoted to R&A (~10%) is the appropriate practice in an era of decreased projected funding totals. However, measuring the effectiveness of the R&A programs, and their role in generating future space missions, is beyond the scope of this work. It will be just as important to consider the funding to continue operating the current set of satellites and whether maintaining that is of higher value to the taxpayer and whether the current fraction of the money spent developing missions is appropriate in an epoch of decreased expectations. Ultimately, scientific questions and not strictly political and budgetary considerations will help guide NASA in making these choices.

The lack of dramatic of scientific trends exhibited in the proposal pressure figures over this period is not terribly surprising. While science marches forward, its pace is often lurching with long periods of relative stability offset with occasional bursts of new activity. This constancy can be inferred from the information in Table 5 or Figures 3-6. Where there are dramatic increases in specific research areas we can often account for them with changes in programmatic directions as opposed to scientific focus shifts. For example, the increased proposal pressure in the 2009 ASP research area for Star Formation and Pre-Main Sequence Stars arose from *Spitzer Space Telescope* archival data becoming available through the ADP that year. Research into star forming regions is one of *Spitzer's* main scientific strengths. Occasionally new research areas are identified for inclusion in a specific program. Such is the case with Atomic and Molecular databases (ADP) or Dark Energy and Cosmic Microwave Background (ATP). Therefore measurements of any variation in these areas will need to wait for further proposal cycles. Within the ATP there may be a decrease in the number of proposals in the area of Interstellar Medium and Galactic Structure. The average fraction of proposal received for this area is $12\pm1\%$ but the number of proposals received has dropped from the mid to upper 20's per year to the low 20's by the decade end. However, it is also important to note that the precise definitions given in the annual call for proposals for the individual research areas may differ slightly from year to year thereby causing shifts of a particular scientific topic from one research area to another. For example, in the 2008 ROSES call the ATP research area *Normal Stars and Star Formation* was used for brown dwarf studies while they are included in the *Stellar Astrophysics* research area in the 2010 ROSES call.

The proposal acceptance rates for any research program are of great importance to the communities they help support. When acceptance rates fall below some value, traditionally thought to be in the 25%-33% range, the perceived 'return-on-investment' of individual researcher's time devoted to proposal preparation is called into question. If acceptance rates become too high, again canonically thought to be greater than ~50%, concerns arise about the quality of funded science. For the three programs under consideration the median acceptance rates were: ADP/LTSA, 29%, (ADP only 32%), APRA, 33%, and ATP, 19%. These figures compare well with recent acceptance rates from the National Science Foundation where the global acceptance rate for all programs was 25% and 29% for the Mathematical and Physical Science programs specifically (Mervis 2009) prior to the 2009 stimulus funding increase. For astronomy and astrophysics the acceptance rate for the years 2006 through 2008 was 22% (N. Sharp, private communication).

The ATP program has traditionally been the most oversubscribed research program in the NASA Astrophysics Division headquarters managed portfolio. Not surprisingly, increased support for theoretical research has always been a key recommendation of the National Academy Sciences and NASA advisory bodies. Theoretical work is important both during the mission definition phases through, for example, data challenges to the research community, and in the interpretation of observational results after a mission is operating. The precise boundaries between ADP and ATP can be blurry for this second important function. Attempts at imbedding theory support within flight project lines for developing missions have not met with uniform success. One of the primary difficulties associated with increasing the level of theory support at NASA has been the lack of quantifiable metric for what the appropriate absolute or relative level of support actually should be. The 30% of a grants program suggested by the *Astronomy and Astrophysics in the New Millennium* Panel on Theory, Computation and Data Exploration has not been achieved within the R&A program. Currently ATP funding stands at about 20% of the total.

The balance of science areas funded by the ADP, APRA and ATP is a topic of annual discussion with the Astrophysics Subcommittee of the NASA Advisory Council. Scientists working within a specific research area are often of the opinion that their discipline or area is underfunded or not receiving the amount of funding that it did in the past. Table 5 and Figures 3-6 show that the fraction of the selected program is essentially equal to the proposal pressure for any given research area. Graphically, the figures provide a quick insight into any trends for individual research areas. Were the programs perfectly 'balanced' all vectors would be horizontal. A reasonable expectation would be for vectors to be randomly distributed, upward and downward, reflecting random fluctuations in proposal strength and review panel preferences with time. Two possible reasons might explain why there are areas which experience more proposal pressure than the selected fraction, *i.e.*, where vectors consistently are lower on the left (*e.g.*, ADP: AGNs, APRA: Laboratory Astrophysics): there exists a large community of proposers relative to what NASA chooses to support, or a community that consistently underperforms in its proposal writing relative to other disciplines. The former explanation is much more plausible than the later. For those areas where vectors are consistently lower on the right than the left (*e.g.*, APRA: Suborbital Investigations), NASA is selecting research in those areas at a greater rate than the corresponding proposal pressure would dictate. Again, multiple interpretations are possible: NASA favors this area of research for some reason or proposers in this field are more adept at convincing peer review panels of the scientific importance of their work than the rest of their colleagues. Because NASA takes programmatic considerations into account during it selection process it is natural that some areas may be favored over others.

Over the past roughly 20 years the community has witnessed an explosion in the number of funding proposal opportunities from a variety of sources. Indeed, a frequent concern for scientists, particularly those supporting themselves on grant funding alone, is the large demand on their time imposed by the sheer number of separate opportunities to propose for funding. Over the 2001-2009 period the average number of proposals submitted per unique principal investigator to each of the programs was: ADP 2.0, APRA 2.7, ATP 2.6. This averages out to about one proposal per investigator every 4 years for the ADP and one proposal every 3 years for APRA and ATP. By itself these proposal submission rates would not seem to be excessively burdensome, but this does not count the numerous other opportunities to propose such as those offered by the individual missions, NOAO, NRAO, and the NSF. Globally these three NASA programs received $462\pm67$ proposals per year and accepted $127\pm26$. The removal of the LTSA program after 2004 and temporary suspension of the ADP in 2005 makes it difficult to determine whether the proposal rate stayed constant or changed significantly during the period.

Proposal reviewers are instructed to evaluate the proposed budgets for cost realism. NASA relies on the experience of its panelists to assess the necessity and accuracy of proposed work and associated expenses. At the same time, NASA clearly instructs reviewers that the absolute value of the dollars proposed should not be a factor in their assessment of the proposed costs. Some efforts are inherently more expensive than others. Ultimately, NASA programmatic balancing discussions influence matters related to the absolute costs of the proposals selected. These data show that proposed costs do not inject an inherent bias in the peer review process, nor do they significantly skew subsequent programmatic choices that must take into account the cost of proposed efforts. Scientific merit of the proposal carries the majority of the weight for NASA selection of astrophysics proposals in these programs. The relatively narrow spread in the proposed (and selected) costs within a given program is not surprising. The ROSES solicitation for these programs always furnishes a prediction of the funds available and likely number of selections. Armed with this information proposers naturally target their budgets around an average cost per effort.

The average single year proposed costs given in Table 9 compare reasonably with the analogous single year grant average from the National Science Foundation's Astronomy Division. There, the canonical faculty-level grant averages approximately $120K/year (N. Sharp, private communication). ATP grants are of similar value while ADP grants typically representing smaller efforts are slightly lower in costs. APRA grants that include hardware development or sub-orbital payloads are naturally larger, in this case by about a factor of three. The distribution of funding among these three programs reflects priorities derived from both the scientific community and NASA itself.

These data contain a nearly decade long measure of NASA astrophysics R&A proposal selection and funding choices that are not specifically tied to a flight missions. They may prove useful to officials as measures of program status, or as starting points should future changes in research program balance be warranted. I made no attempt to assess the appropriateness or effectiveness of any of the research funding associated with the programs here. Such measures would require additional data such as number of published papers, citation rates, number of Ph.D. theses or patents generated, for example. An evaluation of the NASA astrophysics research programs for its effectiveness remains an interesting, but unexplored arena.

The author thanks Heather Lancaster of the NASA Peer Review Services for her assistance with compiling data from the NSPIRES database.

## 5. References
Mervis, J. 2009, *Science*, 326, 1181.

**Table 1**

Proposal Data Fields

Unique proposal identifier (generated by the proposal ingestion system)

Principal Investigator (PI) first and last name

PI institution linked organization (from Central Contractor Registration database)

PI host institution name

Research Area/Topic Category (listed in §1.1 above for each program)

Flag for accepted or declined proposal

Proposed budgets for each year

Awarded budgets for each year

Program Name

Proposal Year

Research Title

Research Summary

Discipline area (APRA proposals only)

Principal Investigator (PI) or Co-Investigator (CoI) proposal flag (APRA Suborbital Investigations proposals only)

**Table 2**

Submitted and Accepted ADP/LTSA Proposals

| Research Area | | 2001 | 2002 | 2003 | 2004 | 2005 | 2006 | 2007 | 2008 | 2009 | Median |
|---|---|---|---|---|---|---|---|---|---|---|---|
| Star Formation & Pre-Main Sequence Stars | | 10 | 15 | 14 | 14 | 0 | 5 | 5 | 8 | 24 | 12 |
| | Accepted | 3 | 5 | 2 | 4 | 0 | 0 | 4 | 4 | 12 | 4 |
| Stellar Astrophysics | | 7 | 9 | 16 | 5 | 0 | 7 | 6 | 3 | 11 | 7 |
| | Accepted | 3 | 2 | 5 | 1 | 0 | 3 | 5 | 1 | 5 | 3 |
| Post-Main Sequence Stars & Collapsed Objects | | 35 | 37 | 33 | 26 | 0 | 19 | 24 | 12 | 22 | 25 |
| | Accepted | 4 | 7 | 8 | 7 | 0 | 6 | 13 | 5 | 10 | 7 |
| Binary Systems | | 29 | 20 | 12 | 9 | 0 | 6 | 8 | 8 | 6 | 8.5 |
| | Accepted | 7 | 7 | 3 | 2 | 0 | 1 | 5 | 4 | 2 | 3.5 |
| Interstellar Medium & Galactic Structure | | 30 | 29 | 26 | 25 | 0 | 14 | 16 | 11 | 24 | 24.5 |
| | Accepted | 6 | 7 | 7 | 7 | 0 | 5 | 8 | 3 | 10 | 7 |
| Normal Galaxies | | 32 | 27 | 19 | 23 | 0 | 11 | 13 | 20 | 25 | 21.5 |
| | Accepted | 4 | 8 | 5 | 6 | 0 | 6 | 4 | 7 | 11 | 6 |
| Active Galaxies and Quasars | | 29 | 43 | 33 | 27 | 0 | 16 | 13 | 12 | 24 | 25.5 |
| | Accepted | 4 | 8 | 7 | 6 | 0 | 5 | 6 | 3 | 12 | 6 |
| Large Scale Cosmic Structures | | 36 | 45 | 48 | 37 | 0 | 22 | 14 | 16 | 24 | 30 |
| | Accepted | 6 | 8 | 10 | 9 | 0 | 9 | 4 | 5 | 9 | 8.5 |
| Atomic/Molecular Databases | | 0 | 0 | 1 | 0 | 0 | 0 | 0 | 5 | 3 | 0 |
| | Accepted | 0 | 0 | 0 | 0 | 0 | 0 | 0 | 3 | 2 | 0 |
| Total Received | | 208 | 225 | 202 | 166 | 0 | 100 | 99 | 95 | 163 | 154 |
| Total Accepted | | 37 | 52 | 47 | 42 | 0 | 35 | 49 | 35 | 73 | 45 |
| Success Fraction | | 18% | 23% | 23% | 25% | N/A | 35% | 49% | 37% | 45% | 29% |

ADP/LTSA program research areas are listed in the first column. For each year the subsequent columns give the number of submitted [upper value] and accepted or selected [lower value] proposals is given.

The final column lists the median number of submitted and accepted proposals for the research area in column 1.

Table 3a

Submitted and Accepted APRA Research Area Proposals

| APRA Research Area | | 2001 | 2002 | 2003 | 2004 | 2005 | 2006 | 2006.5 | 2007 | 2008 | Median |
|---|---|---|---|---|---|---|---|---|---|---|---|
| Detector Development | | 53 | 59 | 47 | 71 | 62 | 49 | 47 | 44 | 39 | 49 |
| | Accepted | 24 | 23 | 12 | 26 | 11 | 14 | 9 | 8 | 11 | 12 |
| Supporting Technology | | 34 | 40 | 27 | 31 | 25 | 20 | 33 | 28 | 27 | 28 |
| | Accepted | 8 | 8 | 6 | 13 | 6 | 4 | 5 | 8 | 5 | 6 |
| Suborbital Investigations | | 19 | 26 | 22 | 24 | 30 | 38 | 66 | 45 | 37 | 30 |
| | Accepted | 5 | 14 | 19 | 15 | 12 | 11 | 34 | 16 | 12 | 14 |
| Ground-Based Observations | | 5 | 6 | 8 | 7 | 11 | 8 | 6 | 5 | 6 | 6 |
| | Accepted | 1 | 5 | 2 | 1 | 2 | 1 | 1 | 2 | 0 | 1 |
| Laboratory Astrophysics | | 41 | 33 | 25 | 31 | 34 | 28 | 26 | 29 | 28 | 29 |
| | Accepted | 23 | 13 | 7 | 13 | 15 | 9 | 6 | 10 | 8 | 10 |
| Gravitation & Fundamental Physics | | 0 | 10 | 5 | 0 | 0 | 0 | 0 | 0 | 0 | 0 |
| | Accepted | 0 | 6 | 3 | 0 | 0 | 0 | 0 | 0 | 0 | 0 |
| Total Received | | 152 | 174 | 134 | 164 | 162 | 143 | 178 | 151 | 137 | 1395 |
| Total Accepted | | 61 | 69 | 49 | 68 | 46 | 39 | 55 | 44 | 36 | 467 |
| Fraction | | 40% | 40% | 37% | 41% | 28% | 27% | 31% | 29% | 26% | 33% |

APRA program research areas are listed in the first column. For each year the subsequent columns give the number of submitted [upper value] and accepted [lower value] proposals is given. The final column lists the median number of submitted and accepted proposals for the research area in column 1.

Table 3b

Submitted and Accepted APRA Discipline Area Proposals

| APRA Discipline Area | | 2001 | 2002 | 2003 | 2004 | 2005 | 2006 | 2006.5 | 2007 | 2008 | Median |
|---|---|---|---|---|---|---|---|---|---|---|---|
| Cosmic Ray & Particle Astrophysics | | 27 | 24 | 18 | 20 | 12 | 11 | 27 | 18 | 18 | 18 |
| | Accepted | 4 | 8 | 12 | 12 | 4 | 5 | 19 | 8 | 5 | 8 |
| X-ray/Gamma-Ray | | 42 | 61 | 37 | 40 | 35 | 32 | 41 | 49 | 31 | 40 |
| | Accepted | 18 | 25 | 14 | 22 | 11 | 9 | 11 | 16 | 10 | 14 |
| UV/Optical | | 39 | 35 | 30 | 45 | 61 | 50 | 44 | 36 | 33 | 39 |
| | Accepted | 15 | 14 | 8 | 12 | 22 | 12 | 5 | 8 | 6 | 12 |
| IR/Submm | | 36 | 39 | 44 | 52 | 54 | 50 | 65 | 47 | 53 | 50 |
| | Accepted | 19 | 16 | 12 | 19 | 9 | 13 | 20 | 12 | 14 | 14 |
| Other | | 8 | 15 | 5 | 7 | 0 | 0 | 1 | 1 | 2 | 2 |
| | Accepted | 5 | 6 | 3 | 3 | 0 | 0 | 0 | 0 | 1 | 1 |
| Total Received | | 152 | 174 | 134 | 164 | 162 | 143 | 178 | 151 | 137 | 1395 |
| Total Accepted | | 61 | 69 | 49 | 68 | 46 | 39 | 55 | 44 | 36 | 467 |
| Fraction | | 40% | 40% | 37% | 41% | 28% | 27% | 31% | 29% | 26% | 33% |

APRA program discipline areas are listed in the first column. For each year the subsequent columns give the number of submitted [upper value] and accepted [lower value] proposals is given. The final column lists the median number of submitted and accepted proposals for the research area in column 1.

Table 4

Submitted and Accepted ATP Research Area Proposals

| Research Area | | 2001 | 2002 | 2003 | 2004 | 2005 | 2006 | 2007 | 2008 | 2009 | Median |
|---|---|---|---|---|---|---|---|---|---|---|---|
| Dark Energy & CMB | | 0 | 0 | 6 | 0 | 16 | 18 | 20 | 21 | 31 | 16 |
| | accepted | 0 | 0 | 3 | 0 | 2 | 4 | 3 | 2 | 6 | 2 |
| Supernovae & Gamma Ray Bursts | | 0 | 0 | 14 | 9 | 14 | 11 | 15 | 15 | 15 | 14 |
| | accepted | 0 | 0 | 1 | 2 | 2 | 0 | 4 | 3 | 3 | 2 |
| ISM & Galactic Structure | | 26 | 24 | 20 | 29 | 14 | 15 | 17 | 21 | 14 | 20 |
| | accepted | 4 | 4 | 6 | 4 | 2 | 2 | 2 | 4 | 4 | 4 |
| Star Formation & Pre-Main Sequence Stars | | 14 | 17 | 6 | 15 | 14 | 12 | 20 | 14 | 10 | 14 |
| | accepted | 4 | 3 | 3 | 5 | 2 | 2 | 3 | 2 | 1 | 3 |
| Stellar Astrophysics | | 8 | 6 | 8 | 7 | 6 | 7 | 0 | 8 | 16 | 7 |
| | accepted | 0 | 1 | 0 | 1 | 1 | 1 | 0 | 3 | 2 | 1 |
| Gravitational Astrophysics and Fundamental Physics | | 5 | 12 | 16 | 0 | 22 | 26 | 27 | 25 | 17 | 17 |
| | accepted | 0 | 3 | 4 | 0 | 5 | 7 | 9 | 8 | 4 | 4 |
| Collapsed Objects & X-ray Astrophysics | | 30 | 36 | 15 | 23 | 28 | 29 | 23 | 17 | 27 | 27 |
| | accepted | 7 | 8 | 4 | 5 | 4 | 6 | 6 | 2 | 7 | 6 |
| Large Scale Cosmic Structure | | 33 | 26 | 19 | 29 | 23 | 19 | 21 | 13 | 36 | 23 |
| | accepted | 8 | 5 | 5 | 4 | 3 | 2 | 5 | 2 | 4 | 4 |
| Normal Galaxies | | 9 | 13 | 8 | 13 | 12 | 13 | 12 | 13 | 7 | 12 |
| | accepted | 2 | 2 | 1 | 1 | 3 | 1 | 2 | 2 | 1 | 2 |
| Active Galaxies and AGNs | | 10 | 10 | 18 | 0 | 10 | 18 | 25 | 21 | 22 | 18 |
| | accepted | 4 | 1 | 5 | 0 | 0 | 5 | 4 | 4 | 3 | 4 |
| Other | | 0 | 0 | 3 | 0 | 14 | 7 | 3 | 9 | 5 | 3 |
| | accepted | 0 | 0 | 0 | 0 | 4 | 2 | 0 | 2 | 1 | 0 |
| Total Received | | 135 | 144 | 133 | 125 | 173 | 175 | 183 | 177 | 200 | 1445 |
| Total Accepted | | 29 | 27 | 32 | 22 | 28 | 32 | 38 | 34 | 36 | 278 |
| Success Fraction | | 21% | 19% | 24% | 18% | 16% | 18% | 21% | 19% | 18% | 19% |

ATP program research areas are listed in the first column. For each year the subsequent columns give the number of submitted [upper value] and accepted [lower value] proposals is given. The final column lists the median number of submitted and accepted proposals for the research area in column

Table 5
Comparison of proposed and accepted fractions

| Research Area | Proposal Submission Fraction | | Acceptance Fraction | |
|---|---:|---|---:|---|
| **ADP** | | | | |
| Star Formation & Pre-Main Sequence Stars | 7% | ±6.3% | 9% | ±11.6% |
| Stellar Astrophysics | 7% | ±3.3% | 9% | ±5.5% |
| Post-Main Sequence Stars and Collapsed Objects | 19% | ±5.0% | 16% | ±11.3% |
| Binary Systems | 8% | ±6.3% | 10% | ±6.7% |
| Interstellar Medium and Galactic Structure | 16% | ±4.5% | 12% | ±8.7% |
| Normal Galaxies | 11% | ±6.6% | 10% | ±10.4% |
| Active Galaxies and Quasars | 13% | ±6.6% | 15% | ±9.1% |
| Large Scale Cosmic Structures | 16% | ±4.5% | 12% | ±8.3% |
| Atomic/Molecular Databases | 4% | ±1.9% | 7% | ±3.9% |
| **ATP** | | | | |
| Dark Energy and Cosmic Microwave Background | 9% | ±4.7% | 6% | ±6.5% |
| Supernovae and Gamma Ray Bursts | 8% | ±1.5% | 6% | ±4.8% |
| Interstellar Medium and Galactic Structure | 12% | ±1.2% | 13% | ±4.8% |
| Star Formation and Pre-Main Sequence Stars | 8% | ±0.7% | 9% | ±4.1% |
| Stellar Astrophysics | 4% | ±0.7% | 3% | ±3.2% |
| Gravitational Astrophysics and Fundamental Physi | 10% | ±3.4% | 13% | ±10.3% |
| Collapsed Objects and X-ray Astrophysics | 16% | ±1.6% | 19% | ±6.8% |
| Large Scale Cosmic Structure | 13% | ±2.0% | 13% | ±6.2% |
| Normal Galaxies | 7% | ±0.3% | 6% | ±2.5% |
| Active Galaxies and AGNs | 11% | ±2.3% | 13% | ±6.7% |
| Other | 2% | ±0.9% | 0% | N/A |
| **APRA (Research Area)** | | | | |
| Detector Development | 35% | ±8.1% | 28% | ±17.3% |
| Supporting Technology | 20% | ±4.7% | 14% | ±7.1% |
| Suborbital Investigations | 21% | ±10.2% | 33% | ±19.7% |
| Ground-Based Observations | 4% | ±1.4% | 2% | ±3.2% |
| Laboratory Astrophysics | 20% | ±4.2% | 23% | ±12.7% |
| Gravitation and Fundamental Physics | 0% | N/A | 0% | N/A |
| **APRA (Discipline Breakdown)** | | | | |
| Cosmic Ray and Particle Astrophysics | 12% | ±4.1% | 16% | ±10.8% |
| X-ray/Gamma-Ray | 27% | ±7.0% | 29% | ±12.9% |
| UV/Optical | 26% | ±7.2% | 24% | ±11.8% |
| IR/Submm | 34% | ±7.1% | 29% | ±9.8% |
| Other | 1% | ±3.4% | 2% | ±4.8% |

Table 1: Column 1 lists the combination program and research areas considered in the study. Column 2 gives the average fraction of the proposals submitted in the corresponding area. Column 3 gives the average acceptance fraction in the corresponding area.

Table 6
Above Average Success Rate Institutions (ADP), Alphabetical Listing
(ADP average success rate: 29%)

| Institution |
| --- |
| Arizona State University |
| Association of Universities for Research in Astronomy, Inc. |
| Boston University |
| Johns Hopkins University |
| Massachusetts Institute of Technology |
| Michigan State University |
| NASA/Goddard Space Flight Center |
| Naval Research Laboratory |
| Pennsylvania State University |
| Princeton University |
| Smithsonian Institution |
| University of Arizona |
| University of California, Berkeley |
| University of California, Irvine |
| University of Maryland |
| University of Michigan |
| University of Southern California |
| Villanova University |

Table 2: Alphabetical sorting of institutions with above average success rates (> 5 accepted proposals in the 2001-2009 interval)

Table 7
Above Average Success Rate Institutions (APRA), Alphabetical Listing
(APRA average success rate: 33%)

| |
|---|
| California Institute of Technology |
| Columbia University |
| Massachusetts Institute of Technology |
| NASA/Goddard Space Flight Center |
| National Institute of Standards & Technology |
| Ohio State University |
| Pennsylvania State University |
| Princeton University |
| Smithsonian Institution |
| The Washington University |
| University of Arizona |
| University of California, Berkeley |
| University of Chicago |
| University of Maryland |
| University of Michigan |
| University of Wisconsin |

Table 3: I Alphabetical sorting of institutions with above average success rates (> 5 accepted proposals in the 2001-2009 interval)

Table 8
Above Average Success Rate Institutions (ATP), Alphabetical Listing
(ATP average success rate: 19%)

| |
|---|
| California Institute of Technology |
| Columbia University |
| Cornell University |
| Drexel University |
| Harvard University |
| NASA/Goddard Space Flight Center |
| New York University |
| Northwestern University |
| Ohio State University |
| Princeton University |
| University of California, Berkeley |
| University of California, San Diego |
| University of California, Santa Cruz |
| University of Colorado |
| University of Illinois |
| University of Texas at Austin |
| University of Virginia |

Table 4: Alphabetical sorting of institutions with above average success rates (> 5 accepted proposals in the 2001-2009 interval)

Table 9

Proposed Costs ($FY10) Comparisons for Accepted and Non-selected Proposals

| Program | Total Cost (Accepted) | Total Cost (Not Accepted) | K-S Statistic $D$ | Probability |
|---|---|---|---|---|
| ADP Total | $205,214 | $202,789 | 0.069 | 0.295 |
| ADP Year 1 | $84,744 | $85,161 | 0.070 | 0.151 |
| APRA Total | $1,103,340 | $1,026,270 | 0.104 | 0.002 |
| APRA Year 1 | $335,875 | $307,928 | 0.102 | 0.003 |
| APRA Total (no suborbital) | $855,206 | $733,593 | 0.093 | 0.037 |
| APRA Year 1 (no suborbital) | $282,320 | $249,467 | 0.087 | 0.063 |
| ATP Total | $426,784 | $388,822 | 0.102 | 0.017 |
| ATP Year 1 | $139,511 | $129,105 | 0.078 | 0.130 |

Table 5: Column 1 lists the program element and the budget considered in the comparison, either the entire budget over the life of the proposed work, or just the first year budget. Columns 2 and 3 list the average costs for the accepted and rejected proposals. Column 4 gives the 2-dimensional Kolmogorov-Smirnov statistic $D$ for the two populations in columns 2 and 3. Column 5 gives the probability that the distributions are drawn from identical distributions. Values less than 0.01 are significant at better than 99% confidence.

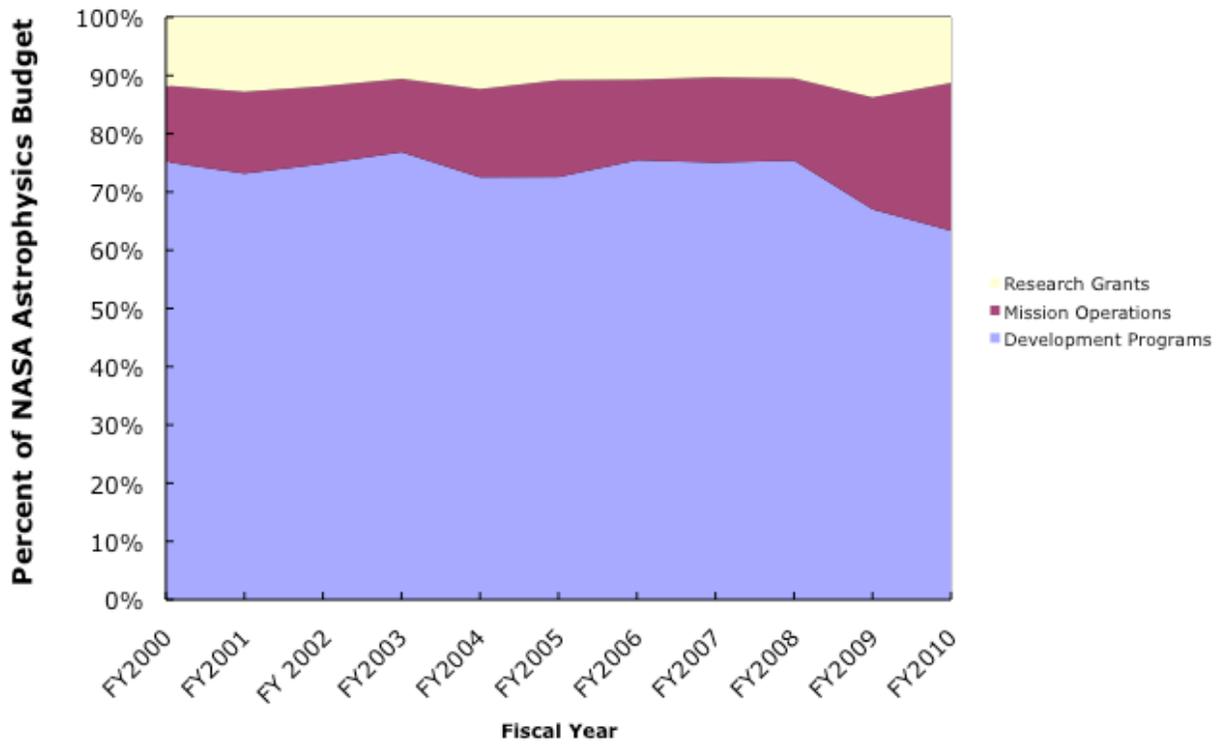

Figure 1: Relative fractions of NASA Astrophysics funding devoted to missions in development (blue), missions in their operations phase (maroon), and research grants (cream) as a function of government fiscal year.

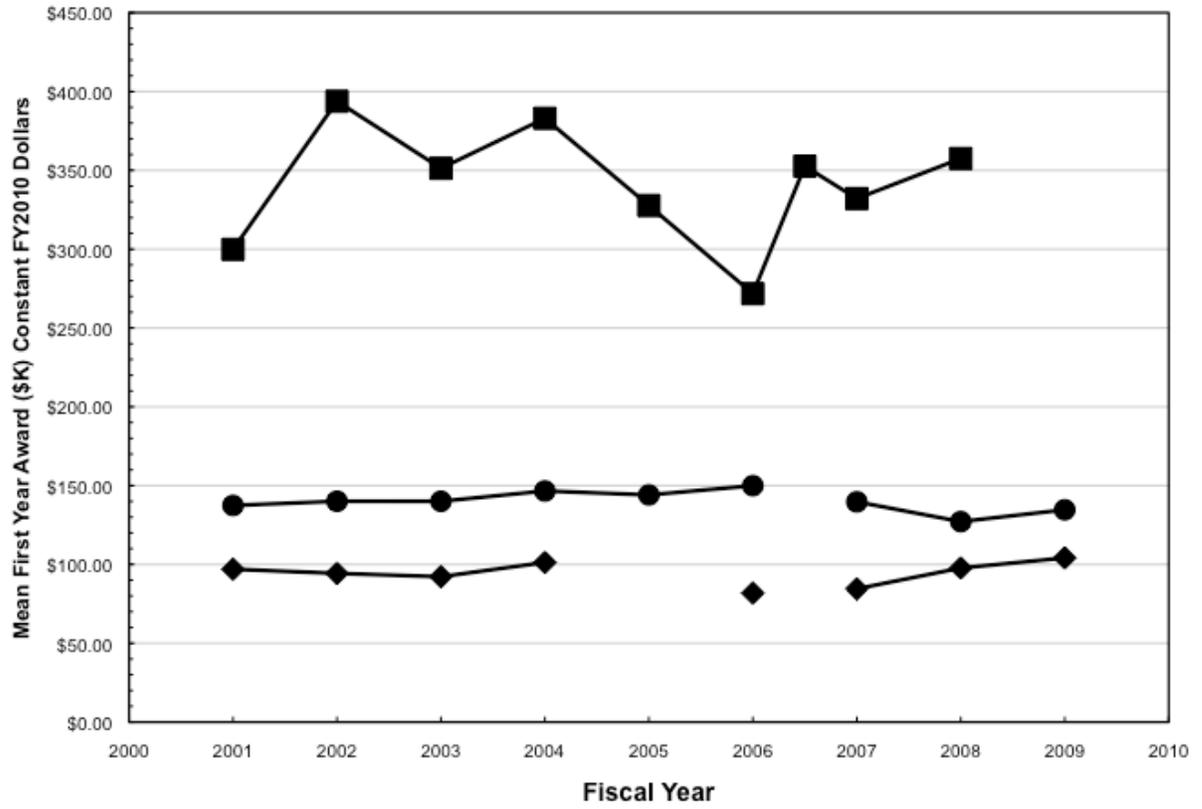

Figure 2: Funding awarded ($M, adjusted to FY10 dollars) in the first year of newly accepted grants for the APRA (squares), ADP (triangles) and ATP (diamonds) programs.

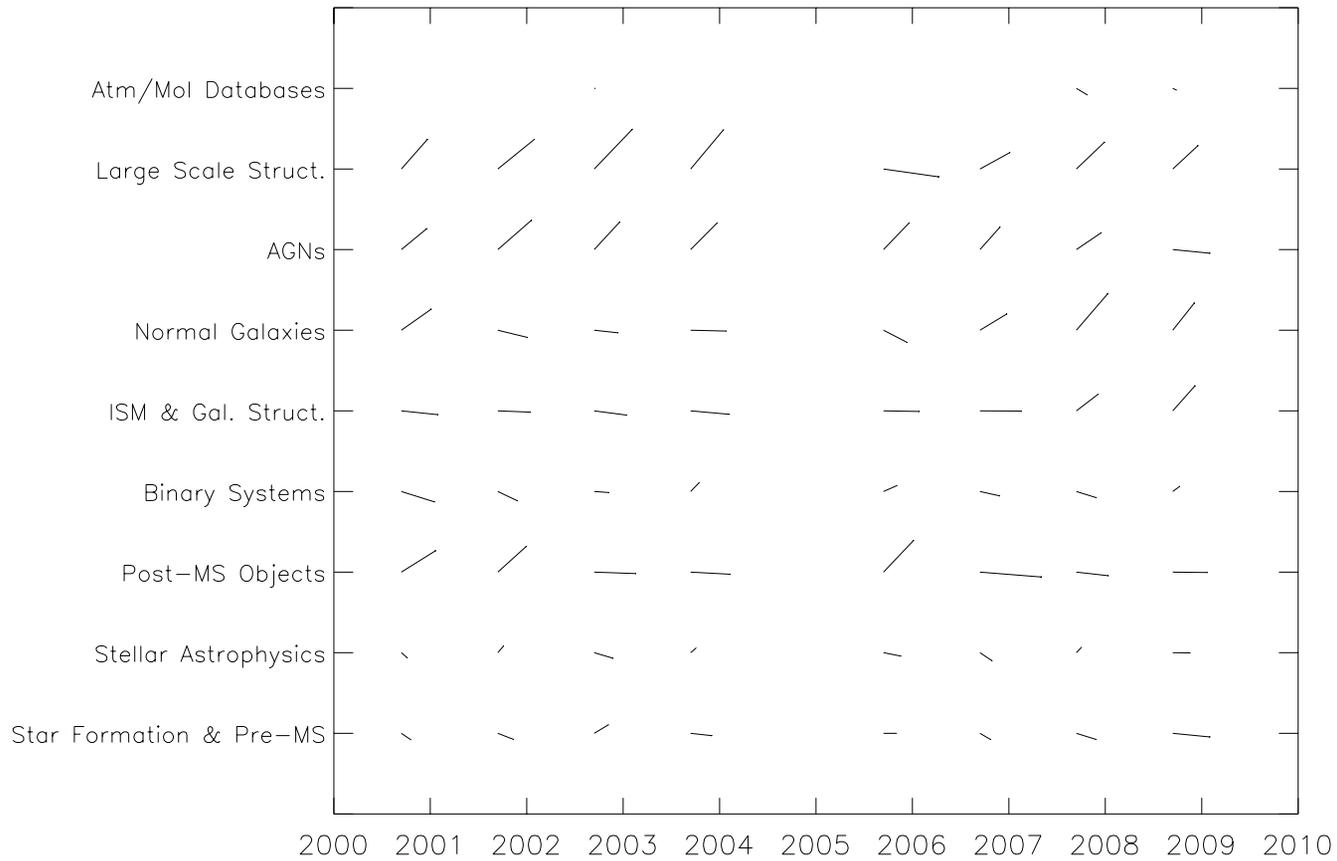
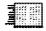

Figure 3: Graphical depiction of the relative balance between proposal pressure and selected program fraction for the research areas of the ADP program listed in §1.1.1. Section 3.2 describes the method for constructing the image from the yearly data. Vectors lower on the left indicate greater proposal pressure than selected fraction.

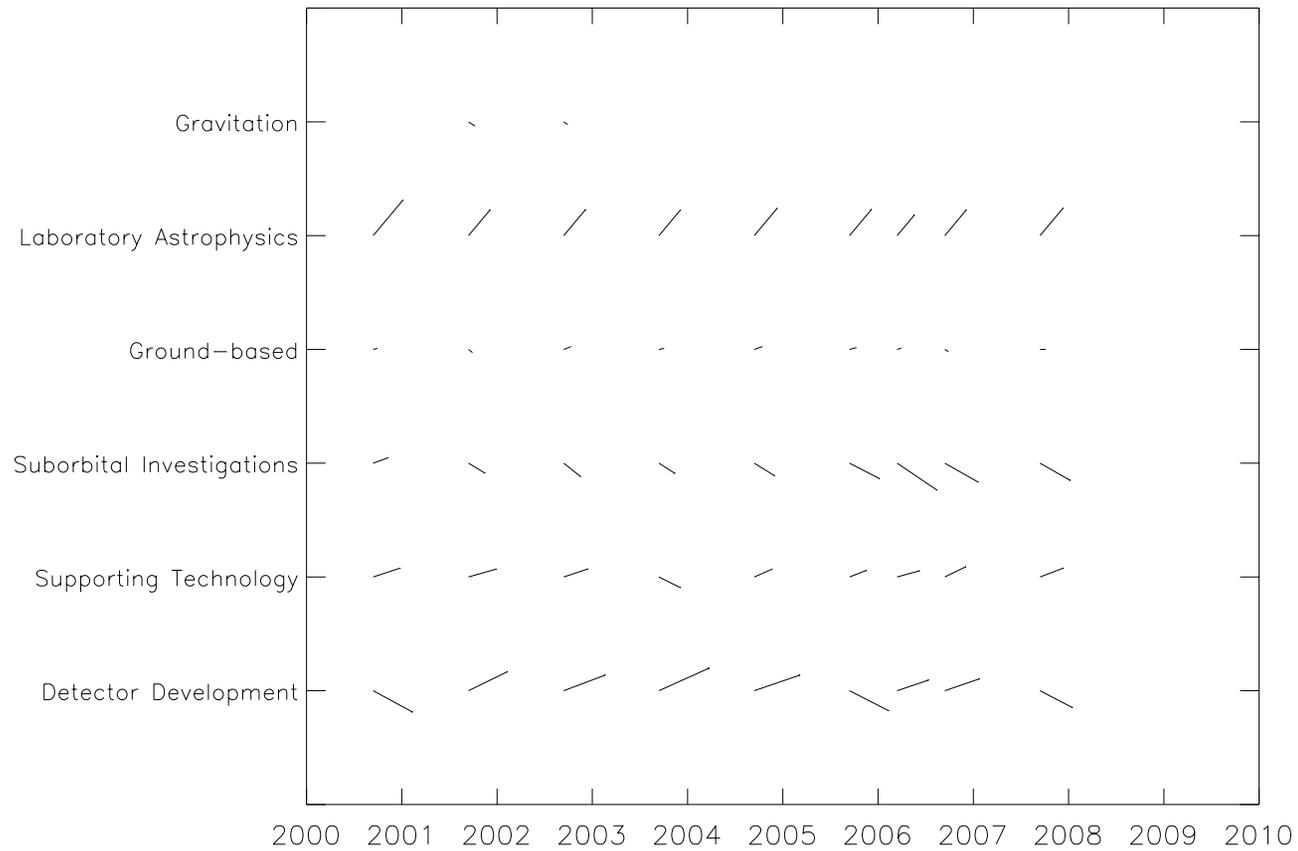

Figure 4: Graphical depiction of the relative balance between proposal pressure and selected program fraction for the research areas of the APRA program listed in §1.1.1. Section 3.2 describes the method for constructing the image from the yearly data. Vectors lower on the left indicate greater proposal pressure than selected fraction.

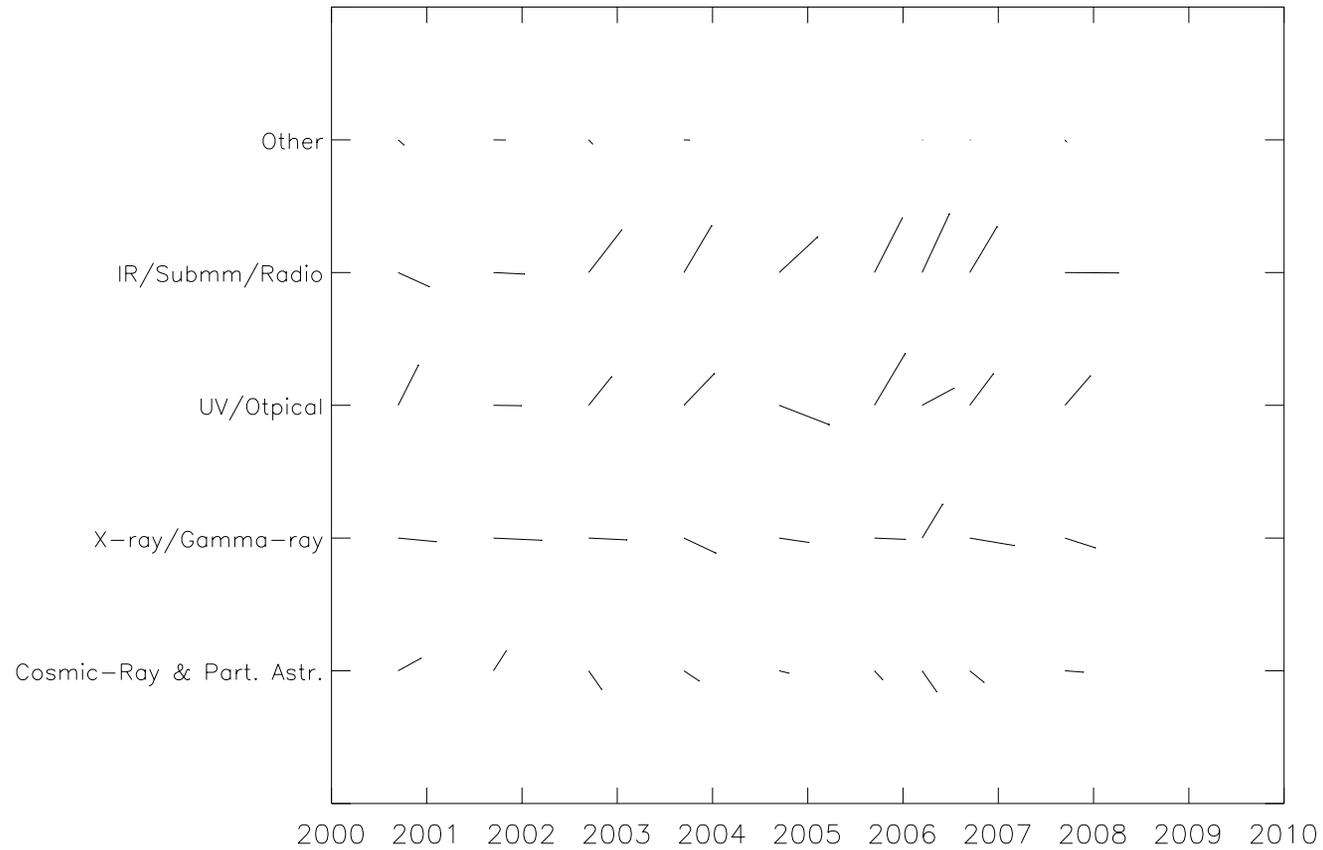

Figure 5: Graphical depiction of the relative balance between proposal pressure and selected program fraction for the discipline areas of the APRA program listed in §1.1.1. Section 3.2 describes the method for constructing the image from the yearly data. Vectors lower on the left indicate greater proposal pressure than selected fraction.

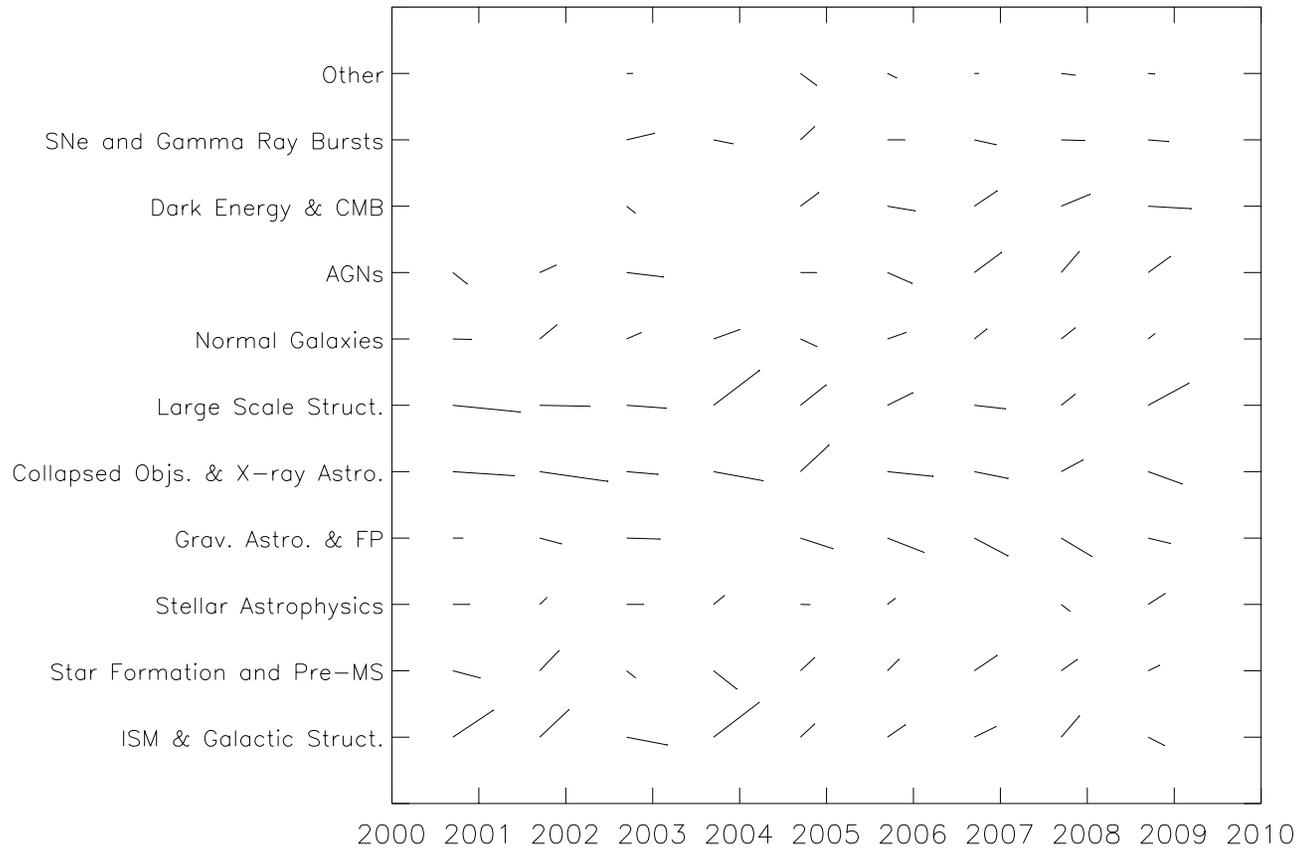
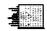

Figure 6: Graphical depiction of the relative balance between proposal pressure and selected program fraction for the research areas of the ATP program listed in §1.1.1. Section 3.2 describes the method for constructing the image from the yearly data. Vectors lower on the left indicate greater proposal pressure than selected fraction.

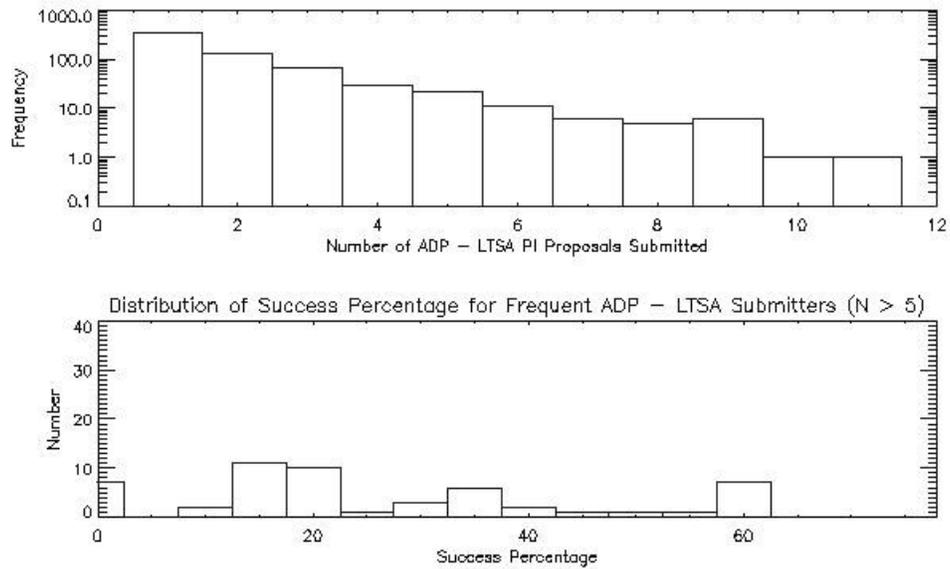
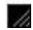

Figure 7: Upper panel: The distribution of the number of ADP proposals submitted by individual principal investigators over the 2001-2009 period. Lower panel: The distribution of the success rate for frequent submitters (more than 1 proposal every other year on average) over the same period.

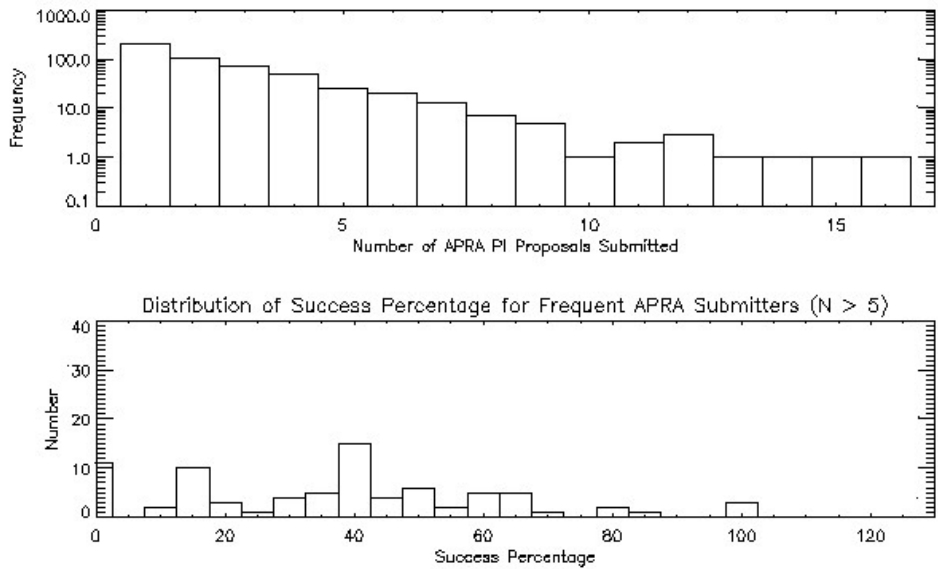
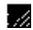

Figure 8: Upper panel: The distribution of the number of APRA proposals submitted by individual principal investigators over the 2001-2009 period. Lower panel: The distribution of the success rate for frequent submitters (more than 1 proposal every other year on average) over the same period.

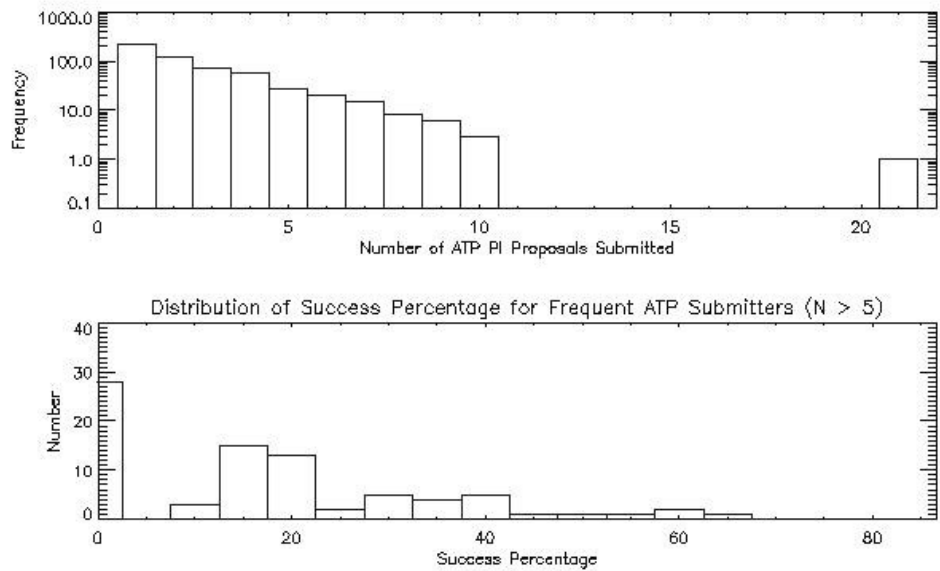
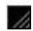

Figure 9: Upper panel: The distribution of the number of ATP proposals submitted by individual principal investigators over the 2001-2009 period. Lower panel: The distribution of the success rate for frequent submitters (more than 1 proposal every other year on average) over the same period

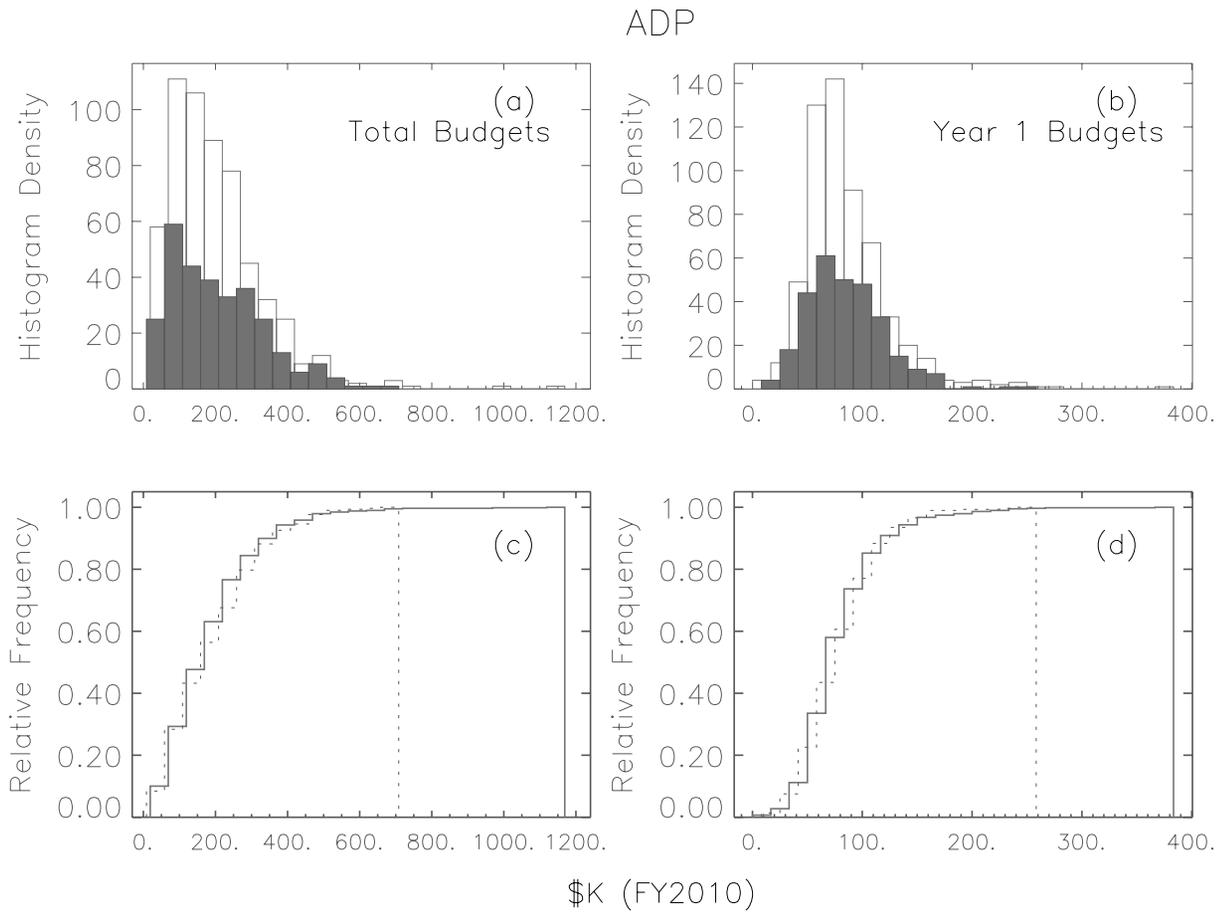

Figure 10: (a) Distribution of total budgets for all ADP proposals with accepted proposals denoted by filled bars and rejected proposals by open bars. (b) Distribution of Year 1 budgets for all ADP proposals with accepted proposals denoted by filled bars and rejected proposals by open bars. (c) Cumulative histogram for total budgets for accepted [solid line] and rejected [dotted line]. (d) Cumulative histogram for Year 1 budgets for accepted [solid line] and rejected [dotted line]. All dollar amounts converted to Fiscal Year 2010 for this comparison.

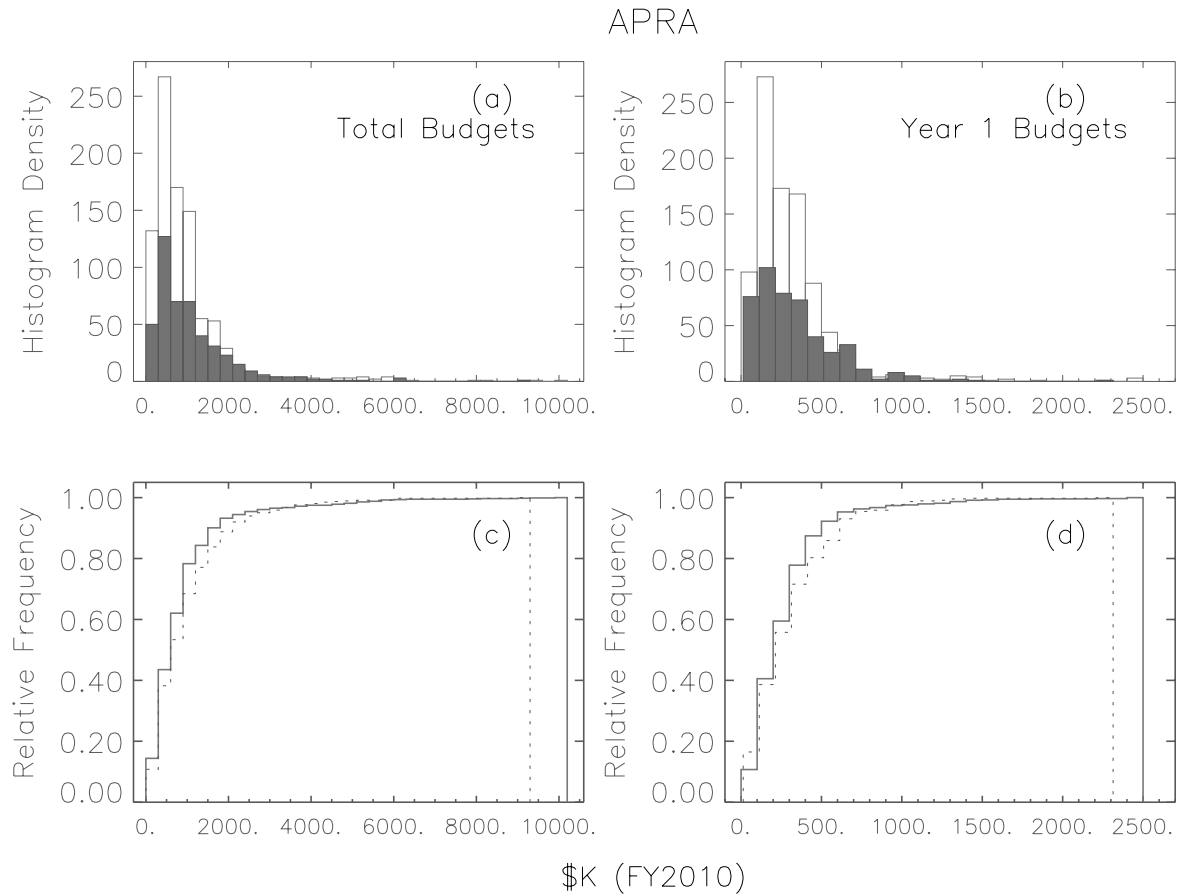

Figure 11: (a) Distribution of total budgets for all APRA proposals with accepted proposals denoted by filled bars and rejected proposals by open bars. (b) Distribution of Year 1 budgets for all APRA proposals with accepted proposals denoted by filled bars and rejected proposals by open bars. (c) Cumulative histogram for total budgets for accepted [solid line] and rejected [dotted line]. (d) Cumulative histogram for Year 1 budgets for accepted [solid line] and rejected [dotted line]. All dollar amounts converted to Fiscal Year 2010 for this comparison.

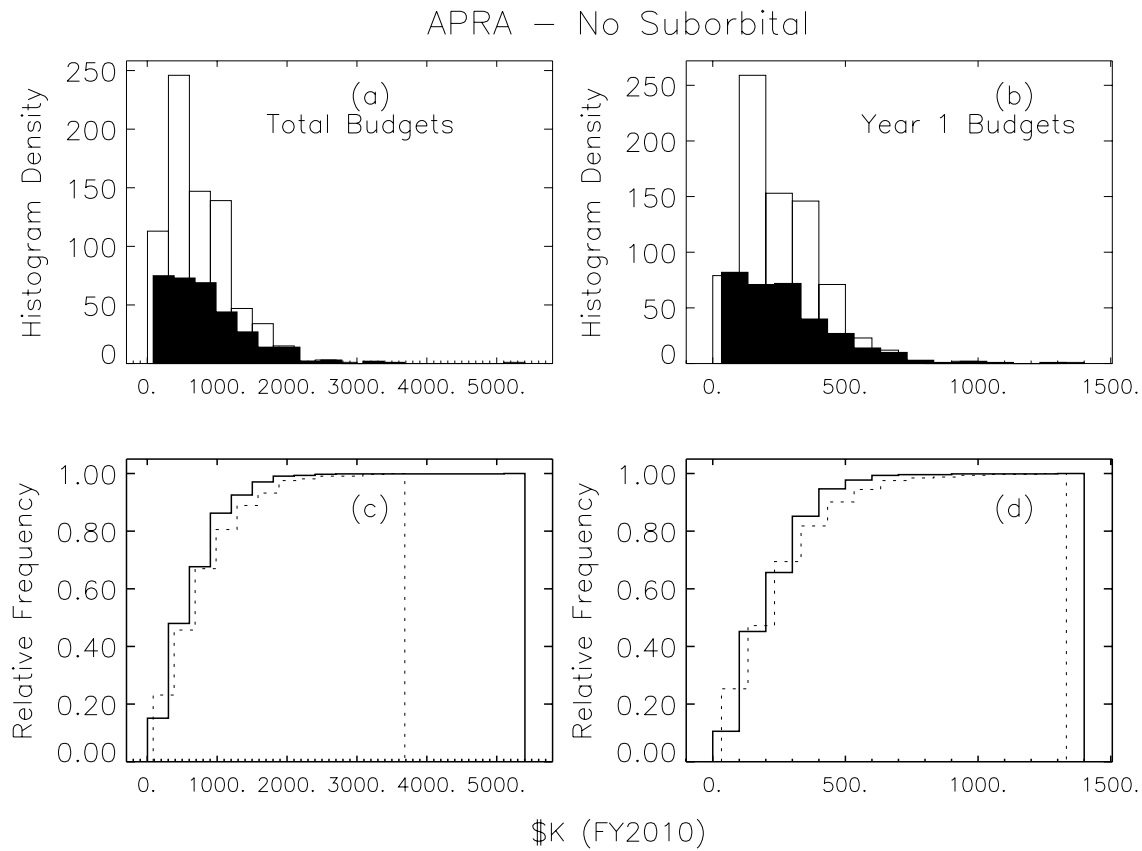

Figure 12: (a) Distribution of total budgets for all APRA proposals, but excluding all suborbital investigations with accepted proposals denoted by filled bars and rejected proposals by open bars. (b) Distribution of Year 1 budgets for all APRA proposals with accepted proposals denoted by filled bars and rejected proposals by open bars. (c) Cumulative histogram for total budgets for accepted [solid line] and rejected [dotted line]. (d) Cumulative histogram for Year 1 budgets for accepted [solid line] and rejected [dotted line]. All dollar amounts converted to Fiscal Year 2010 for this comparison.

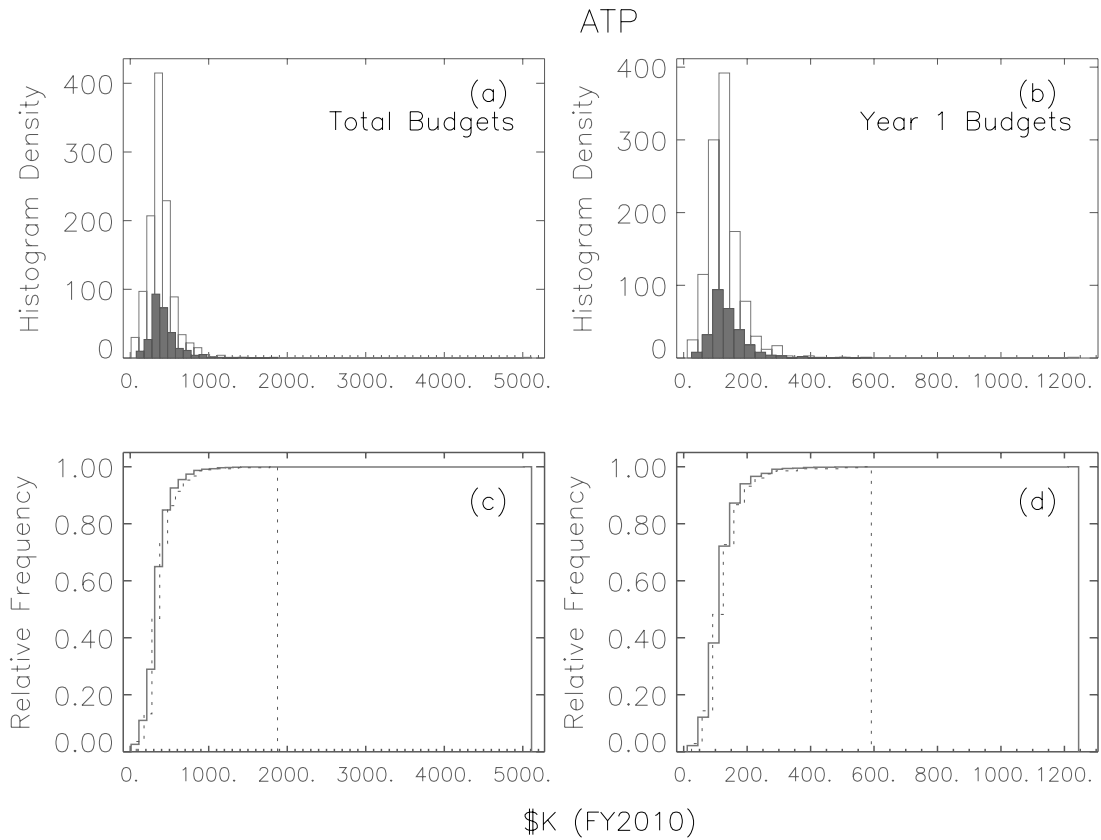

Figure 13: (a) Distribution of total budgets for all ATP proposals with accepted proposals denoted by filled bars and rejected proposals by open bars. (b) Distribution of Year 1 budgets for all ATP proposals with accepted proposals denoted by filled bars and rejected proposals by open bars. (c) Cumulative histogram for total budgets for accepted [solid line] and rejected [dotted line]. (d) Cumulative histogram for Year 1 budgets for accepted [solid line] and rejected [dotted line]. All dollar amounts converted to Fiscal Year 2010 for this comparison.